\documentclass[aps,prb,twocolumn,superscriptaddress,showpacs]{revtex4}

\usepackage{graphicx}
\usepackage{dcolumn}
\usepackage{bm}

\begin{document}

\title{Comment on ``Low-temperature phonon thermal conductivity of single-crystalline 
Nd$_2$CuO$_4$: Effects of sample size and surface roughness" by  
Li {\it et al.}}

\author{X. F. Sun}

\affiliation{Hefei National Laboratory for Physical Sciences at 
Microscale, University of Science and Technology of China, Hefei, 
Anhui 230026, P. R. China}

\author{Yoichi Ando}
\affiliation{Institute of Scientific and Industrial Research, 
Osaka University, Ibaraki, Osaka 567-0047, Japan}

\date{\today}

\begin{abstract}

In this Comment, we show that the experimental data reported by 
Li {\it et al.} (Phys. Rev. B {\bf 77}, 134501 (2008)) do not 
support the phonon specular reflection at low temperatures for 
high-$T_c$ cuprates, because the phonon mean free path $\ell$ 
calculated from their thermal conductivity data is much smaller 
than the averaged sample width above $\sim$ 100 mK for both 
Nd$_2$CuO$_4$ and YBa$_2$Cu$_3$O$_{6.0}$ single crystals.

\end{abstract}


\maketitle

In a recent paper, Li {\it et al.}\cite{Li} reported their 
low-temperature thermal conductivity ($\kappa$) studies on 
Nd$_2$CuO$_4$ single crystals. Based on their examinations of the 
effect of sample size and surface roughness on the phonon heat 
conductivity $\kappa_p$, they concluded the dominance of specular 
phonon reflection in high-$T_c$ cuprates and proposed an 
empirical power law $T^{\alpha}$ ($\alpha <$ 2) to describe 
$\kappa_p/T$. They further used this empirical formula to 
re-analyze our published data\cite{Sun_nonuniversal} for Zn-doped 
YBa$_2$Cu$_3$O$_y$ (YBCO) and claimed that the universal thermal 
conductivity of $d$-wave quasiparticles at $T \to$ 0 was 
confirmed. In this comment, we show that both their experiments 
and data analysis are problematic or self-inconsistent, and 
therefore their conclusion on our data of Zn-doped YBCO lacks 
reasonable ground.

\begin{figure}
\includegraphics[clip,width=7cm]{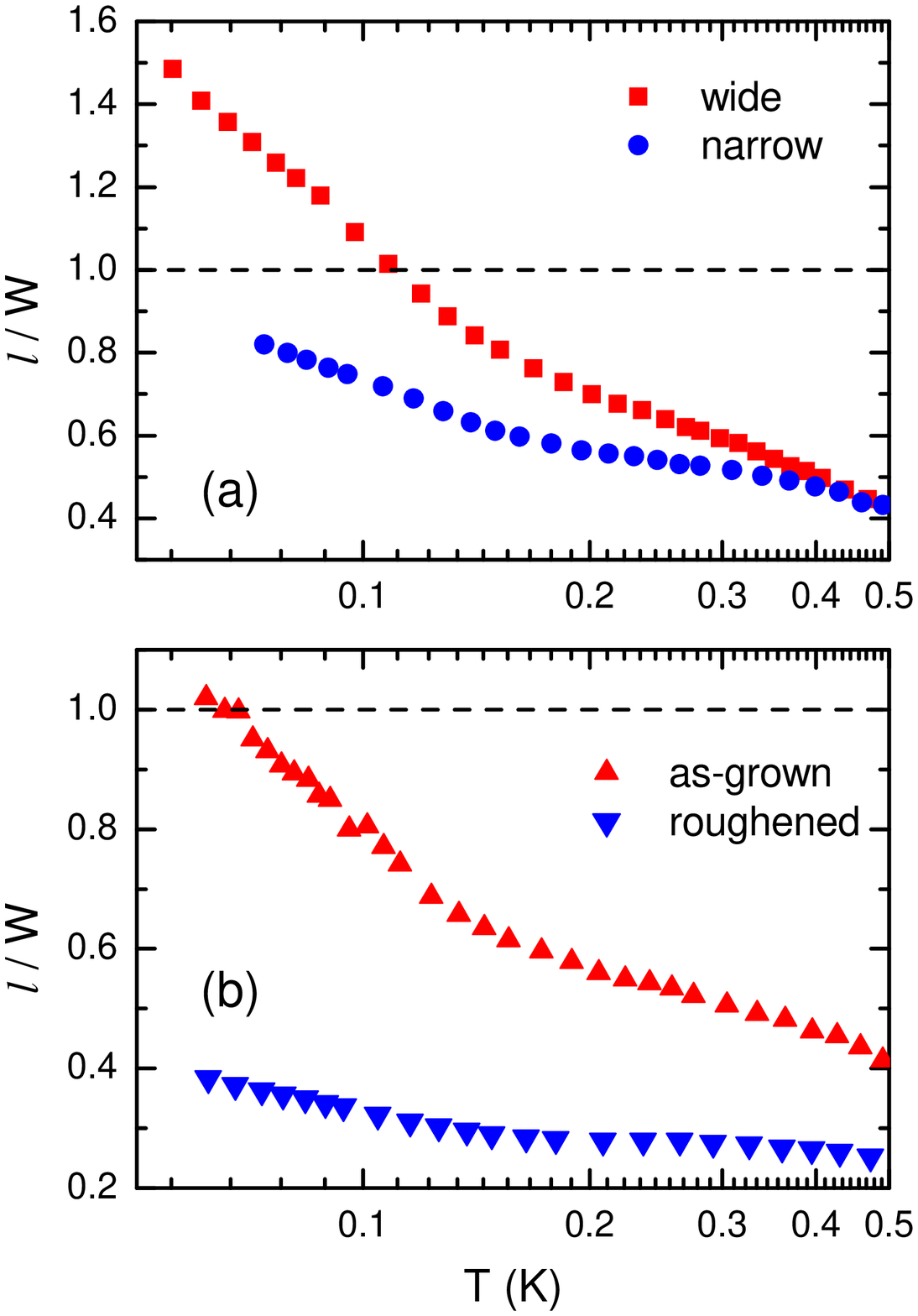}
\caption{(color online) Temperature dependence of the calculated 
phonon mean free path $\ell$ divided by the averaged sample width 
$W$ for Nd$_2$CuO$_4$. The raw data of $\kappa$ are taken from 
Ref. \onlinecite{Li}. }
\end{figure}

Li {\it et al.}\cite{Li} showed that the thermal conductivities 
of a wide Nd$_2$CuO$_4$ sample and a narrow one (cut from the 
wide sample) differ, as did the thermal conductivities of an 
as-grown sample and the same sample after surface roughening, and 
claimed that the observed differences support their assertion 
that phonons are in the boundary scattering regime below 0.5 K 
and that the specular reflection off sample surface is important, 
causing the $T^{\alpha+1}$ ($\alpha <$ 2) dependence of $\kappa$ 
in a rather broad temperature range. Let us first point out that, 
contrary to what Li {\it et al.} claimed, the phonons are {\it 
not} really specularly reflected in their samples by correctly 
calculating the phonon mean free path $\ell$ from their data and 
comparing it with the averaged sample width $W$, which is usually 
taken to be $2/\sqrt{\pi}$ times the geometrical mean width 
$\bar{w}$.\cite{Taillefer} Note that previous studies on the 
specular reflection unanimously found\cite{Hurst,Pohl,Thacher} 
that $\ell$ gets much longer than $W$ when the specular 
reflection becomes important. Actually, since $\kappa_p$ is equal 
to $\frac{1}{3}C\bar{v}\ell$ where $C = \beta T^3$ is the phonon 
specific heat and $\bar{v}$ is the averaged sound velocity, and 
both $\beta$ and $\bar{v}$ are known experimentally for 
Nd$_2$CuO$_4$,\cite{Li_NCO, Specific_heat, Velocity} one can 
easily calculate $\ell$ in Li {\it et al.}'s samples.\cite{Li} 
The coefficient $\beta$ of Nd$_2$CuO$_4$ were reported to be 
about 0.42--0.50 mJ/mol$\cdot$K.\cite{Specific_heat, note1} To 
avoid underestimating the value of $\ell$, we choose $\beta$ = 
0.42 mJ/mol$\cdot$K for the calculation. Another parameter, the 
averaged sound velocity, can be calculated by $\bar{v} = \sum_i 
{v_i}^{-2}/\sum_i {v_i}^{-3}$,\cite{Casimir} where $i$ represents 
the three acoustic phonon modes. Here, we use the experimental 
values of the in-plane sound velocities of Nd$_2$CuO$_4$ from 
Ref. \onlinecite{Velocity}, $v_L$ = 6050 m/s, $v_{T1}$ = 4220 m/s 
and $v_{T2}$ = 2460 m/s (which are also used by Li {\it et 
al.}\cite{Li}), to obtain $\bar{v}$ = 2940 m/s. Figure 1 shows 
the temperature dependence of $\ell$ calculated from the data in 
Ref. \onlinecite{Li}. One can see that for most cases $\ell$ only 
becomes comparable to $W$ at the lowest temperature and it stays 
much smaller than $W$ in the temperature range where Li {\it et 
al.} argued\cite{Li} that the boundary scattering is taking place 
(below 0.5 K). For surface-roughened sample, $\ell$ is so small 
that it is much smaller than $W$ even at the lowest temperature.  
Therefore, the magnitude of $\ell$ is obviously inconsistent with 
the assertion that phonons are boundary scattered below 0.5 K in 
Nd$_2$CuO$_4$. In fact, since $\kappa_p = \frac{1}{3}(\beta 
T^3)\bar{v}\ell$, a departure of $\kappa_p$ from the $T^3$ 
dependence always comes from a $T$ dependence of $\ell$, and it 
occurs either when the specular reflection is important (which 
causes $\ell > W$) or when phonons are not in the boundary 
scattering regime (in which case $\ell < W$). Therefore, one can 
conclude that in Nd$_2$CuO$_4$ at temperatures higher than 
$\sim$100 mK, phonons are {\it not} in the boundary scattering 
regime and the phenomenological $T^{\alpha+1}$ ($\alpha <$ 2) 
dependence comes from some additional scattering that only {\it 
reduces} $\ell$.\cite{Berman} 

Furthermore, one can see that a comparison of the thermal 
conductivity data between the wide sample and the narrow one 
indicates some self-inconsistency if one tries to employ the 
specular reflection to explain the $T^{\alpha+1}$ dependence: 
When the sample is gradually cooled, one expects that $\ell$ 
gradually increases when the microscopic scattering is becoming 
less effective and finally $\ell$ is equal to the averaged width, 
where the phonons enter the boundary scattering limit (about 0.5 
K in Li {\it et al.}'s speculation). Upon further cooling, $\ell$ 
would become larger than the averaged width if the specular 
reflection is at work. Naturally, the characteristic temperature 
where $\ell$ is becoming comparable or equal to the averaged 
width should not strongly decrease with the sample size; rather, 
it should move to higher temperature upon decreasing the width. 
This is clearly opposite to what Li {\it et al.}'s data imply, as 
is demonstrated in Fig. 1(a). Another self-inconsistency appeared 
when Li {\it et al.} compared the data between the as-grown 
sample and the surface-roughened one,\cite{Li} that is, the two 
data sets differ by as much as a factor of 2 at 0.5 K where 
$\kappa$ should already be mostly governed by microscopic 
scattering mechanisms, which should give a 
sample-size-independent thermal conductivity. (Here, we are not 
talking about thermal {\it conductance}.) This is clearly 
inconsistent with what they claimed.\cite{Li} Note that the 
cutting or sanding process could bring some cracks into the 
samples, and our own experience taught us that small cracks can 
easily be created when flux-grown crystals are cut; the narrower 
and thinner the sample, the more significantly the cracks affect 
the transport results.

It should be noted that Li {\it et al.} showed an estimation of 
$\ell$ for the surfaced-roughed sample by using the following 
formula,\cite{Li}
\begin{equation}
 \kappa_p = \frac{2}{15} \pi^2k_B \big(\frac{k_BT}{\hbar} \big)^3 <v^{-2}>\ell 
 \label{kappa_p},
\end{equation}
in which only one parameter $<v^{-2}>$ is necessary for the 
calculation. To obtain this formula, there is an approximation 
for the phonon specific heat coefficient
\begin{equation}
 \beta = \frac{2\pi^2}{5} \big(\frac{k_B^4}{\hbar^3} \big)^3 <\frac{1}{v^3}> 
 \label{beta},
\end{equation}
where $<\frac{1}{v^3}>$ is the average of the inverse third power 
of the long-wavelength phase velocities of the three acoustic 
modes.\cite{Casimir, Ashcroft_Mermin} Calculating $\beta$ with 
this approximation using measured sound velocities would be valid 
if the sound velocity solely reflects the dispersion of the 
acoustic phonon near the $\Gamma$ point. However, 
Eq.(\ref{kappa_p}) is apparently not precise when one compares 
the experimental results of $\beta$ and the sound 
velocities:\cite{Specific_heat, Velocity} Using the experimental 
velocities from Ref. \onlinecite{Velocity}, one obtains $\beta$ = 
0.20 mJ/mol$\cdot$K from formula (\ref{beta}), which is smaller 
than the experimental value of $\beta$ (0.42--0.50 
mJ/mol$\cdot$K).\cite{Specific_heat} In this sense, Li {\it et 
al.}'s calculation using formula (\ref{kappa_p}) is not well 
supported by the experimental literature. Furthermore, their 
calculation\cite{Li} itself is problematic and $\ell$ is at least 
30\% overestimated: they used a roughly averaged value of the 
velocities ($\bar{v}$ = 4000 m/s) for calculating $<v^{-2}>$ 
through 1/$\bar{v}^2$ (= 6.25 $\times$ $10^{-8}$ s$^2$/m$^2$); 
however, one can easily notice that $<v^{-2}>$ should be the 
averaged value of the inverse second power of the sound 
velocities of the three acoustic modes, which is $<v^{-2}>$ = 
8.29 $\times$ $10^{-8}$ s$^2$/m$^2$. 

\begin{figure}
\includegraphics[clip,width=7cm]{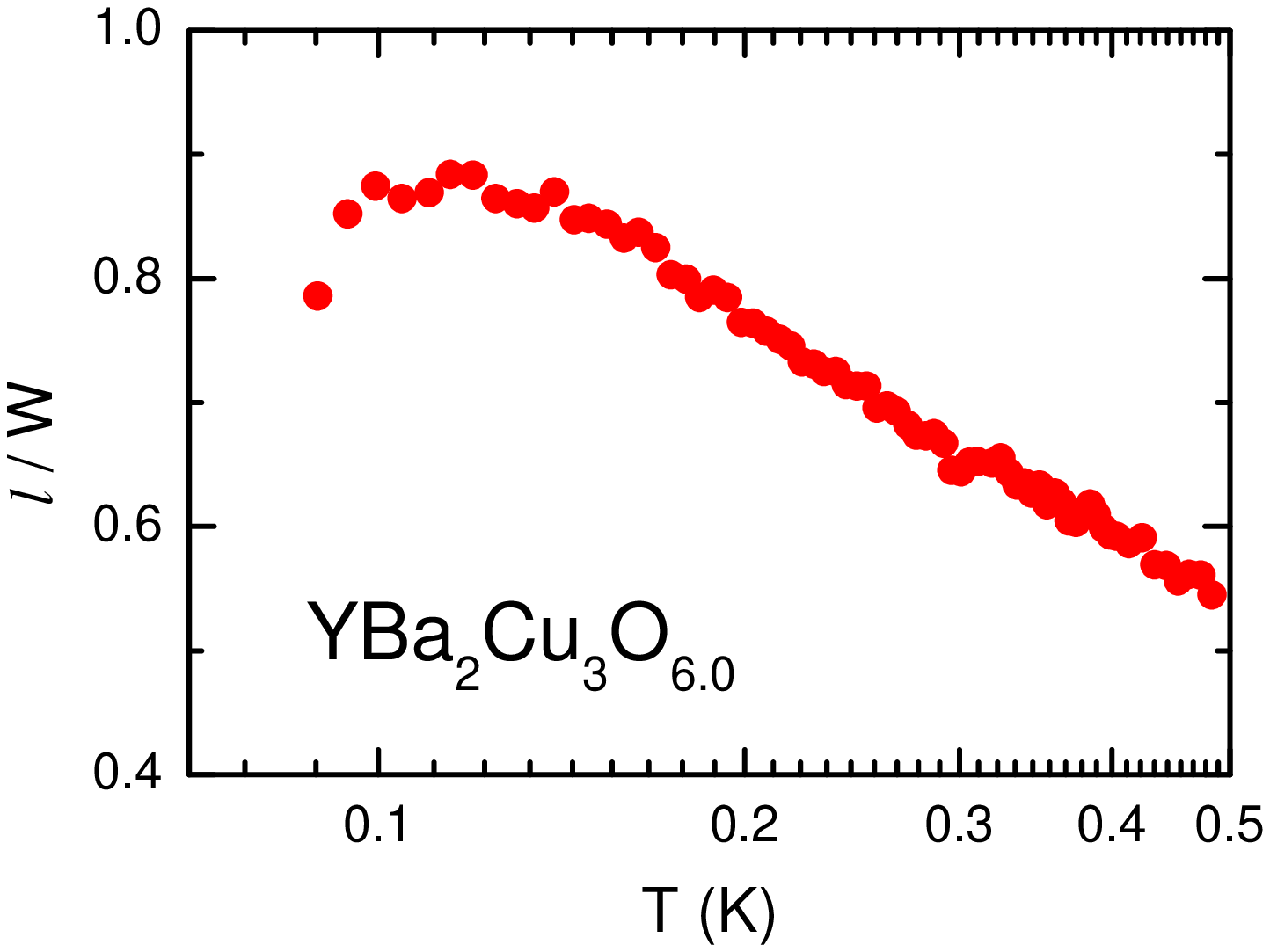}
\caption{(color online) Temperature dependence of the calculated 
phonon mean free path $\ell$ divided by the averaged sample width 
$W$ for YBa$_2$Cu$_3$O$_{6.0}$. The raw data of $\kappa$ are 
taken from Ref. \onlinecite{Taillefer}.}
\end{figure}

Therefore, Li {\it et al.}'s data\cite{Li} for Nd$_2$CuO$_4$ do 
not really support their speculation that the specular phonon 
reflection is important in cuprates. Then, what about another 
cuprate, YBCO, which they simply assumed to have the same phonon 
transport properties as Nd$_2$CuO$_4$? One can easily find the 
answer from the thermal conductivity data of an insulating YBCO 
sample with $y$ = 6.0, reported by the same 
group.\cite{Taillefer} In Fig. 2, we show the temperature 
dependence of $\ell$ in this sample where the heat transport is 
purely phononic, using the data from Ref. \onlinecite{Taillefer}. 
It is clear that the YBCO crystal shows a standard phonon 
transport behavior, with the boundary scattering limit only 
achieved below 140 mK, where $\ell$ is close to the averaged size 
and $\kappa$ shows a $T^3$ dependence. Actually, it was already 
demonstrated\cite{Taillefer} that the low-$T$ $\kappa/T$ data of 
this sample below 140 mK can be reasonably well fit to $a + bT^2$ 
with zero intercept ($a$ = 0).

Neglecting those circumstances,  Li {\it et al.}\cite{Li} used 
the formula $\kappa/T = a + bT^{\alpha}$ to re-fit our low-$T$ 
thermal conductivity data of the superconducting YBCO and 
Zn-doped YBCO [Ref. \onlinecite{Sun_nonuniversal}] and concluded 
that $\kappa_0/T$ is unchanged upon Zn doping. Based on such an 
analysis, they claimed that our data confirm the universal 
thermal conductivity of $d$-wave quasiparticles for $T \to$ 0. 
However, it must be pointed out that even if the phonon heat 
conductivity showed a $T^{\alpha+1}$ dependence, the fitting to 
$\kappa/T = a + bT^{\alpha}$ is meaningful {\it only when the 
{\it electronic} heat conductivity is a linear function of 
temperature}. In this regard, one should not forget an important 
fact, which was first reported by the Taillefer group,\cite{Hill} 
that the electronic contribution to the heat transport in clean 
YBCO increases rapidly with temperature following a $T^3$ law, 
due to the thermal creation of heat-carrying quasiparticles near 
the nodes. If one takes into account this $T^3$ dependence of the 
electronic term at $T \neq$ 0, one can easily see that the 
analysis of our YBCO data by Li {\it et al.} is obviously 
erroneous; in contrast, our original 
fitting\cite{Sun_nonuniversal} with $\kappa/T = a + bT^2$ is 
robust against the appearance of the $T^3$ electronic 
contribution and, hence, is reliable for extracting the residual 
thermal conductivity. In fact, the ``main effect" of Zn doping to 
cause a dramatic suppression of the {\it slope} of $\kappa/T$, 
which Li {\it et al.} noted in Ref. \onlinecite{Li}, is mostly 
due to the diminishment of this electronic $T^3$ term.

It would also be useful to note that Nd$_2$CuO$_4$ used in Li 
{\it el al.}'s work is not an ideal system for studying the 
purely phononic heat transport because of its complicated 
magnetism.\cite{theory} In fact, theoretically the Nd magnon 
modes are predicted to be effectively gapless in zero 
field,\cite{theory} so the conjecture of Li {\it et al.} that the 
Nd magnons are gapped and do not contribute to the heat transport 
does not have any theoretical support. If the magnon contribution 
is not negligible in Nd$_2$CuO$_4$, the true mean free path of 
phonons would be even smaller than those shown in Fig. 1.

\begin{acknowledgments}

This work was supported by the National Natural Science 
Foundation of China (10774137 and 50721061), the National Basic 
Research Program of China (2006CB922005), and KAKENHI Grant No. 
19674002.

\end{acknowledgments}


\begin{thebibliography}{}

\bibitem{Li}
S. Y. Li, J.-B. Bonnemaison, A. Payeur, P. Fournier, C. H. Wang, 
X. H. Chen, and L. Taillefer, Phys. Rev. B {\bf 77}, 134501 
(2008).

\bibitem{Sun_nonuniversal}
X. F. Sun, S. Ono, Y. Abe, S. Komiya, K. Segawa, and Y. Ando, 
Phys. Rev. Lett. {\bf 96}, 017008 (2006).

\bibitem{Hurst}
W. S. Hurst and D. R. Frankl, Phys. Rev. {\bf 186}, 801 (1969). 

\bibitem{Pohl}
R. O. Pohl and B. Stritzker, Phys. Rev. B {\bf 25}, 3608 (1982). 

\bibitem{Thacher}
P. D. Thacher, Phys. Rev. {\bf 156}, 975 (1967). 

\bibitem{Taillefer}
L. Taillefer, B. Lussier, R. Gagnon, K. Behnia, and H. Aubin, 
Phys. Rev. Lett. {\bf 79}, 483 (1997).

\bibitem{Li_NCO}
S. Y. Li, L. Taillefer, C. H. Wang, and X. H. Chen, Phys. Rev. 
Lett. {\bf 95}, 156603 (2005).

\bibitem{Specific_heat}
M. F. Hundley, J. D. Thompson, S.-W. Cheong, Z. Fisk, and S. B. 
Oseroff, Physica C {\bf 158}, 102 (1989); N. T. Hien, V. H. M. 
Duijn, J. H. P. Colpa, J. J. M. Franse, and A. A. Menovsky, Phys. 
Rev. B {\bf 57}, 5906 (1998); R. Kuentzler, G. Pourroy, A. 
Tigheza, and Y. Dossmann, Solid State Commun. {\bf 78}, 113 
(1991).

\bibitem{Velocity}
D. V. Fil', I. G. Kolobov, V. D. Fil', S. N. Barilo, and D. I. 
Zhigunov, Low Temp. Phys. {\bf 21}, 937 (1995).

\bibitem{note1}
In Nd$_2$CuO$_4$, one cannot use very low-temperature specific 
heat data for analyzing the phonon term because of the appearance 
of a magnetic peak, so in Ref. \onlinecite{Specific_heat} 
the $\beta$ value was obtained from the simple fitting 
of $T^3$ to the data at intermediate temperature. It is known 
that in the temperature range 0.02 $< T / \theta_D <$ 0.1 
($\theta_D$ is the Debye temperature), one 
had better use the low-frequency expansion of the Debye function, 
which gives $C_v = \beta T^3 + \beta_5 T^5+\beta_7 T^7 + ...$ 
(See the textbook by A. Tari, {\it Specific Heat of Matter at Low 
Temperatures}, Imperial College Press, 2003). This formula is 
commonly used for analyzing the phonon specific heat at this 
intermediate temperature range. [See, for example, the papers on 
the specific heat of La$_2$CuO$_4$; C. F. Chang {\it et al.}, 
Phys. Rev. Lett. {\bf 84}, 5612 (2000); H.-H. Wen {\it et al.}, 
Phys. Rev. B {\bf 70}, 214505 (2004)]. 
It is important to recognize that the correction 
to the simple $T^3$ analysis can decrease the $\beta$ value, but 
it can never change the $\beta$ value by as much as a factor of 
two, as can be seen in these reference papers. 
More importantly, if the specific heat data 
in the intermediate temperature range can still be 
well fitted to the simple $T^3$ law, which is actually the case 
for Nd$_2$CuO$_4$, it means that the higher power terms are 
effectively negligible in this material. (In Ref. 
\onlinecite{Specific_heat}, the $T^3$ law fits the data well from 
$\sim$ 15 to 25 K or even higher temperature.) This is not 
surprising, because the higher-power term $\beta_5$ is usually 
found to be a small positive value in La$_2$CuO$_4$, but it can even be 
a small negative value in some other materials [e.g., 
see the data for Na$_x$CoO$_2$, Y. Ando {\it et al.}, 
Phys. Rev. B {\bf 60}, 10580 (1999)]. 
Also, it is useful to note that one can neglect the possibility that the
$\beta$ value determined at 15--25 K for Nd$_2$CuO$_4$ might be enhanced
by a magnon contribution, because it is established that in the N\'eel
state of Nd$_2$CuO$_4$, Cu magnons (which are the only possibility at
15--25 K) are gapped, although it is still an open question whether the
Nd magnons (which can only be relevant below 3 K) are gapped as 
well.\cite{theory} 

\bibitem{Casimir}
While the averaged sound velocity $\overline v$ is a rather 
complicated 
average over all acoustic branches in all directions, 
this issue was carefully studied by Casimir and has been well 
sorted out; H. B. G. Casimir, Physica (Amsterdam) {\bf 5}, 495 (1938).

\bibitem{Berman}
R. Berman, {\it Thermal Conduction in Solids} (Oxford University 
Press, Oxford, 1976).

\bibitem{Ashcroft_Mermin}
N. W. Ashcroft and N. D. Mermin, {\it Solid State Physics} 
(Thomson Learing, Inc., 1976).

\bibitem{Hill}
R. W. Hill, C. Lupien, M. Sutherland, E. Boaknin, D. G. Hawthorn, 
C. Proust, F. Ronning, L. Taillefer, R. Liang, D. A. Bonn, and W. 
N. Hardy, Phys. Rev. Lett. {\bf 92}, 027001 (2004).

\bibitem{theory}
R. Sachidanandam, T. Yildirim, A. B. Harris, A. Aharony, and O. 
Entin-Wohlman, Phys. Rev. B {\bf 56}, 260 (1997).


\end{thebibliography}
\end{document}